\documentclass[sigconf]{acmart}
\usepackage{multirow}
\usepackage{booktabs} 
\usepackage{longtable}
\usepackage[T1]{fontenc}
\usepackage{graphicx}    
\usepackage{tabularx}         
\usepackage{array}            
\usepackage{caption}          
\usepackage{booktabs}
\usepackage{multicol} 
\usepackage{arydshln}
\usepackage{xcolor}
\usepackage{colortbl}
\usepackage{pifont} 
\usepackage{xcolor}
\usepackage{pifont} 
\usepackage{algorithm}
\usepackage{algorithmic}
\usepackage{colortbl} 

\definecolor{goodgreen}{RGB}{20, 120, 20} 
\definecolor{badred}{RGB}{180, 20, 20}   
\definecolor{neuorange}{RGB}{200, 100, 0} 

\newcommand{\cmark}{\textcolor{goodgreen}{\ding{51}}} 
\newcommand{\xmark}{\textcolor{badred}{\ding{55}}}   
\newcommand{\good}[1]{\textcolor{goodgreen}{\textbf{#1}}} 
\newcommand{\bad}[1]{\textcolor{badred}{\textbf{#1}}}     
\newcommand{\neu}[1]{\textcolor{neuorange}{\textbf{#1}}}  

\definecolor{studentgray}{HTML}{F2F2F2}
\definecolor{teacherblue}{HTML}{E6F0FF}
\definecolor{analysisyellow}{HTML}{FFFFE0}
\AtBeginDocument{%
  }

\setcopyright{acmlicensed}
\copyrightyear{2018}
\acmYear{2018}
\acmDOI{XXXXXXX.XXXXXXX}
\acmConference[Conference acronym 'XX]{Make sure to enter the correct
  conference title from your rights confirmation email}{June 03--05,
  2018}{Woodstock, NY}
\acmISBN{978-1-4503-XXXX-X/2018/06}
\usepackage[acronym]{glossaries}
\makeglossaries
\newacronym{SFT}{SFT}{Supervised Fine-tuning}
\newacronym{LLM}{LLM}{Large Language Model}
\newacronym{RL}{RL}{Reinforcement Learning}
\newacronym{ERL}{ERL}{Evolutionary Reinforcement Learning}
\newacronym{POMDP}{POMDP}{Partially Observable Markov Decision Process}
\newacronym{SIIP}{SIIP}{Socratic Interdisciplinary Instructional Problem}
\newacronym{MDP}{MDP}{Markov Decision Process}
\newacronym{ZPD}{ZPD}{Zone of Proximal Development}
\newacronym{IRT}{IRT}{Item Response Theory}
\newacronym{EA}{EA}{Evolutionary Algorithm}
\newacronym{SID}{SID}{Socratic Interdisciplinary Dialogues}
\newacronym{HOPE}{HOPE}{Higher-Order Pedagogical Evaluation Score}
\newacronym{PE}{PE}{Prompt Engineering}

\renewcommand\footnotetextcopyrightpermission[1]{} 
\acmConference[]{}{}{} 
\begin{document}

\title{Evolutionary Reinforcement Learning based AI tutor for Socratic Interdisciplinary Instruction}

\author{Mei Jiang}
\email{51265901071@stu.ecnu.edu.cn}
\affiliation{%
  \institution{East China Normal University}
  \city{Shanghai}
  \state{Shanghai}
  \country{China}
}

\author{Haihai Shen}
\email{51285901130@stu.ecnu.edu.cn}
\affiliation{%
  \institution{East China Normal University}
  \city{Shanghai}
  \state{Shanghai}
  \country{China}
}

\author{Zhuo Luo}
\email{10220350421@stu.ecnu.edu.cn}
\affiliation{%
  \institution{East China Normal University}
  \city{Shanghai}
  \state{Shanghai}
  \country{China}
}

\author{Bingdong Li}
\email{bdli@cs.ecnu.edu.cn}
\affiliation{%
  \institution{East China Normal University}
  \city{Shanghai}
  \state{Shanghai}
  \country{China}
}

\author{Wenjing Hong}
\email{hongwj@szu.edu.cn}
\affiliation{%
  \institution{Shenzhen University}
  \city{Shenzhen}
  \state{Guangdong}
  \country{China}
}

\author{Ke Tang}
\email{tangk3@sustech.edu.cn}
\affiliation{%
  \institution{Southern University of Science and Technology}
  \city{Hefei}
  \state{Anhui}
  \country{China}
}

\author{Aimin Zhou}
\email{amzhou@cs.ecnu.edu.cn}
\affiliation{%
  \institution{East China Normal University}
  \city{Shanghai}
  \state{Shanghai}
  \country{China}
}







\renewcommand{\shortauthors}{Trovato et al.}

\begin{abstract}
Cultivating higher-order cognitive abilities—such as knowledge integration, critical thinking, and creativity—in modern STEM education necessitates a pedagogical shift from passive knowledge transmission to active Socratic construction. Although \Glspl{LLM} hold promise for STEM Interdisciplinary education, current methodologies employing \Gls{PE}, \Gls{SFT}, or standard \Gls{RL} often fall short of supporting this paradigm. Existing methods are hindered by three fundamental challenges: the inability to dynamically model latent student cognitive states; severe reward sparsity and delay inherent in long-term educational goals; and a tendency toward policy collapse lacking strategic diversity due to reliance on behavioral cloning. Recognizing the unobservability and dynamic complexity of these interactions, we formalize the \Gls{SIIP} as a structured \Gls{POMDP}, demanding simultaneous global exploration and fine-grained policy refinement. To this end, we propose ERL4SIIP, a novel \Gls{ERL} framework specifically tailored for this domain. ERL4SIIP integrates: (1) a dynamic student simulator grounded in a STEM knowledge graph for latent state modeling; (2) a Hierarchical Reward Mechanism that decomposes long-horizon goals into dense signals; and (3) a LoRA-Division based optimization strategy coupling evolutionary algorithms for population-level global search with PPO for local gradient ascent. Experimental results demonstrate that ERL4SIIP significantly outperforms state-of-the-art baselines in fostering students' higher-order abilities—particularly knowledge integration and interdisciplinary transfer—while exhibiting richer teaching strategy diversity and superior robustness against the "spoon-feeding" trap.
\end{abstract}



\received{20 February 2007}
\received[revised]{12 March 2009}
\received[accepted]{5 June 2009}

\maketitle

\glsresetall
\section{Introduction}
\label{sec:intro}

Fostering higher-order cognitive abilities such as knowledge integration, transfer, critical thinking, and creativity is widely regarded as a central aim of modern STEM education~\cite{osborne2014teaching, vilsmaier2015case}. Research grounded in constructivist theory suggests that these abilities arise through active knowledge construction rather than passive reception, as learners reorganize ideas and resolve cognitive conflict~\cite{chi2009active}. Socratic pedagogy builds on this view by guiding students through purposeful questioning and sustained intellectual challenge, encouraging them to articulate reasoning, confront misconceptions, and gradually refine their understanding~\cite{paul2006thinker, chin2007teacher}.

However, current AI tutors built on \Glspl{LLM} still fall short of translating these pedagogical ideals into practical instructional behavior. Although \Glspl{LLM} possess extensive conceptual knowledge~\cite{bubeck2023sparks}, mainstream alignment approaches—such as \Gls{PE} and \Gls{SFT}—remain fundamentally static. \Gls{SFT} models, constrained by behavioral cloning, are prone to the so-called plot echo effect~\cite{xu2025echoes}, converging toward safe and formulaic average responses rather than offering the adaptive variation needed for personalized scaffolding. More importantly, in the absence of a mechanism for long-horizon planning, these models are implicitly driven to optimize short-term conversational efficiency. As a result, they often default to direct answer delivery rather than sustained reasoning support, making it difficult to elicit genuine cognitive engagement~\cite{kapur2008productive}.

While \Gls{RL} in principle enables optimization beyond behavioral imitation, applying standard \Gls{RL} methods (e.g., PPO) to the \Gls{SIIP} remains fundamentally challenging. First, instructional dialogue inherently forms a \Gls{POMDP} in which student cognition—such as misconceptions, confidence, or frustration—is latent and continuously evolving; collapsing this into a fully observable \Gls{MDP} breaks belief-state tracking and undermines adaptive response generation~\cite{rafferty2016faster}. Second, evidence of conceptual growth is naturally sparse and delayed, so short-term conversational proxies often misrepresent true learning progress and expose models to Goodhart-style reward exploitation, where direct answers become a locally optimal but pedagogically shallow strategy~\cite{skalse2022defining}.
Finally, the immense and compositional action space of natural language renders the optimization landscape highly non-convex, causing gradient-based \Gls{RL} to converge prematurely and collapse into limited behavioral modes, which reduces the strategic diversity needed to support varied learners and sustained reasoning.

To address these deficits, we introduce \textbf{ERL4SIIP}, an \Gls{ERL} framework designed for Socratic Interdisciplinary Instruction. Rather than relying on static imitation, ERL4SIIP enables dynamic, belief-driven interaction by unifying a Dynamic Student Simulator based on a STEM knowledge graph for modeling latent cognition and providing a high-fidelity \Gls{POMDP} environment, a Hierarchical Reward System that decomposes long-horizon instructional goals into dense and non-deceptive feedback to reduce reward hacking, and a LoRA-Division Based Optimization approach that separates exploration from exploitation, using population-level search to maintain diversity while gradient refinement stabilizes learning and prevents strategy collapse.

Our contributions are summarized as follows:
\begin{itemize}
    \item We formalize \Gls{SIIP} as a \Gls{POMDP} and construct a student simulator that explicitly models latent knowledge dynamics. This approach mitigates the unobservability of student cognitive states and establishes a dynamic surrogate environment for pedagogical exploration beyond static datasets.

    \item We introduce a hierarchical reward mechanism to resolve the reward sparsity and hacking challenges in Socratic teaching. This non-deceptive scheme decomposes long-horizon educational goals into dense process signals, ensuring alignment between pedagogical behaviors and student conceptual reorganization. 
    
    \item We propose a LoRA-Division based \Gls{ERL} framework that decouples global exploration from local refinement. By projecting the search into a low-rank manifold, we make population-based evolutionary algorithms computationally feasible for \Glspl{LLM}, effectively preventing the strategy collapse common in standard \Gls{RL}.

\end{itemize}

\section{Related Work}
\label{sec:related_work}

\subsection{\Glspl{LLM} in Education \& \Gls{SFT} Limitations}
Initial efforts leveraged \Gls{PE} to instantiate specific pedagogical tactics, such as refutation or induction, within \Glspl{LLM}~\cite{chang2023prompting, harb2025hitchhiker}. While accessible, these heuristic-based methods lack structural constraints, frequently degenerating into repetitive patterns or hallucinations. To capture more nuanced strategies, recent frameworks like \textit{SocraticLM}~\cite{liu2024socraticlm} and \textit{PlatoLM}~\cite{kong2024platolm} apply \Gls{SFT} to datasets derived from multi-agent simulations. Despite improved coherence, these methods are fundamentally constrained by behavior cloning: they mimic the average expert trajectory within the training distribution but fail to generalize to novel cognitive states or execute the long-horizon planning required for deep conceptual change~\cite{gudibande2023false}.

\subsection{Reinforcement Learning for Educational Alignment}
Recent initiatives leverage \Gls{RL} to optimize pedagogical adaptability. For instance, EduAlign~\cite{song2025cultivating} aligns \Glspl{LLM} with educational principles via multi-dimensional reward modeling, while Pxplore~\cite{wang2025pxplore} optimizes personalized learning paths through a two-stage framework. However, these approaches typically reductively model instruction as a fully observable \Gls{MDP}, neglecting the latent nature of student cognitive states. Moreover, their reliance on sparse outcome signals hinders deep conceptual reorganization and increases vulnerability to reward hacking~\cite{singla2021reinforcement}. We address these structural deficits by formalizing instruction as a \Gls{POMDP} and employing evolutionary strategies to navigate the non-convex optimization landscape of Socratic pedagogy.

\section{Problem Definition}
\subsection{Problem Formulation: SIIP as a POMDP}
\label{sec:problem_def}
We formalize the \Gls{SIIP} as a sequential decision-making process under uncertainty, modeled by a \Gls{POMDP} tuple $\mathcal{P} = \langle \mathcal{S}, \mathcal{A}, \mathcal{T}, \mathcal{R}, \Omega, \mathcal{O}, \gamma \rangle$. Unlike traditional Intelligent Tutoring System (ITS) which assume visible learner states, Socratic pedagogy necessitates continuous inference of the student's evolving cognitive structure.

\noindent Latent State Space ($\mathcal{S}$). The student's unobservable state is a composite vector $s_t = \langle s^{profile}, s^{know}_t, s^{mis}_t, s^{affect}_t \rangle$. It comprises: 
(1) \textit{Static Profile} ($s^{profile}$), invariant parameters encoding the learner's cognitive baseline and personality traits; 
(2) \textit{Knowledge State} ($s^{know}_t$), a continuous vector mapping mastery levels $m \in [0, 1]$ across the domain Knowledge Graph; 
(3) \textit{Misconception State} ($s^{mis}_t$), a discrete vector flagging active erroneous beliefs or rules; and 
(4)\textit{Affective State} ($s^{affect}_t$), tracking real-time dynamics such as frustration and engagement. Detailed definitions are provided in Appendix \ref{app:state_details}.

\noindent Dynamics \& Belief State. The learning process is governed by a stochastic transition function $\mathcal{T}(s' | s, a)$, representing the non-linear evolution of cognition. Since $s_t$ is latent, the agent perceives the environment solely through observations $o_t \in \Omega$ (e.g., student textual responses) and maintains a \textit{Belief State} $b_t(s) = P(s_t = s | h_t)$, updated via Bayesian inference over the interaction history $h_t$.

\noindent Reward Function ($\mathcal{R}$). The scalar signal $r_t = \mathcal{R}(s_t, a_t, s_{t+1})$ evaluates pedagogical efficacy. Crucially, the true reward (cognitive growth) in Socratic contexts is \textit{sparse, delayed, and noisy}. Effective strategies often induce short-term "productive struggle" (low immediate correctness) to achieve long-term conceptual reorganization, complicating direct reward quantification.

\subsection{Optimization Objective}
The goal of the AI tutor is to learn an optimal policy $\pi_{\theta}^*(a_t | b_t)$ that maximizes the expected cumulative discounted reward over the interaction horizon $H$:
\begin{equation}
J(\pi_\theta) = \mathbb{E}_{\tau \sim \pi_\theta} \left[ \sum_{t=0}^{H} \gamma^t r_t \right]
\end{equation}
where $\gamma \in [0, 1]$ is the discount factor balancing immediate feedback and long-term learning outcomes. Solving this optimization problem requires the agent to perform \textit{belief space planning}—strategically choosing actions not just to maximize reward, but to reduce uncertainty about the student's state (information gain).
\section{Methods}

\begin{figure*}[htbp]
    \centering
    \includegraphics[width=0.95\textwidth]{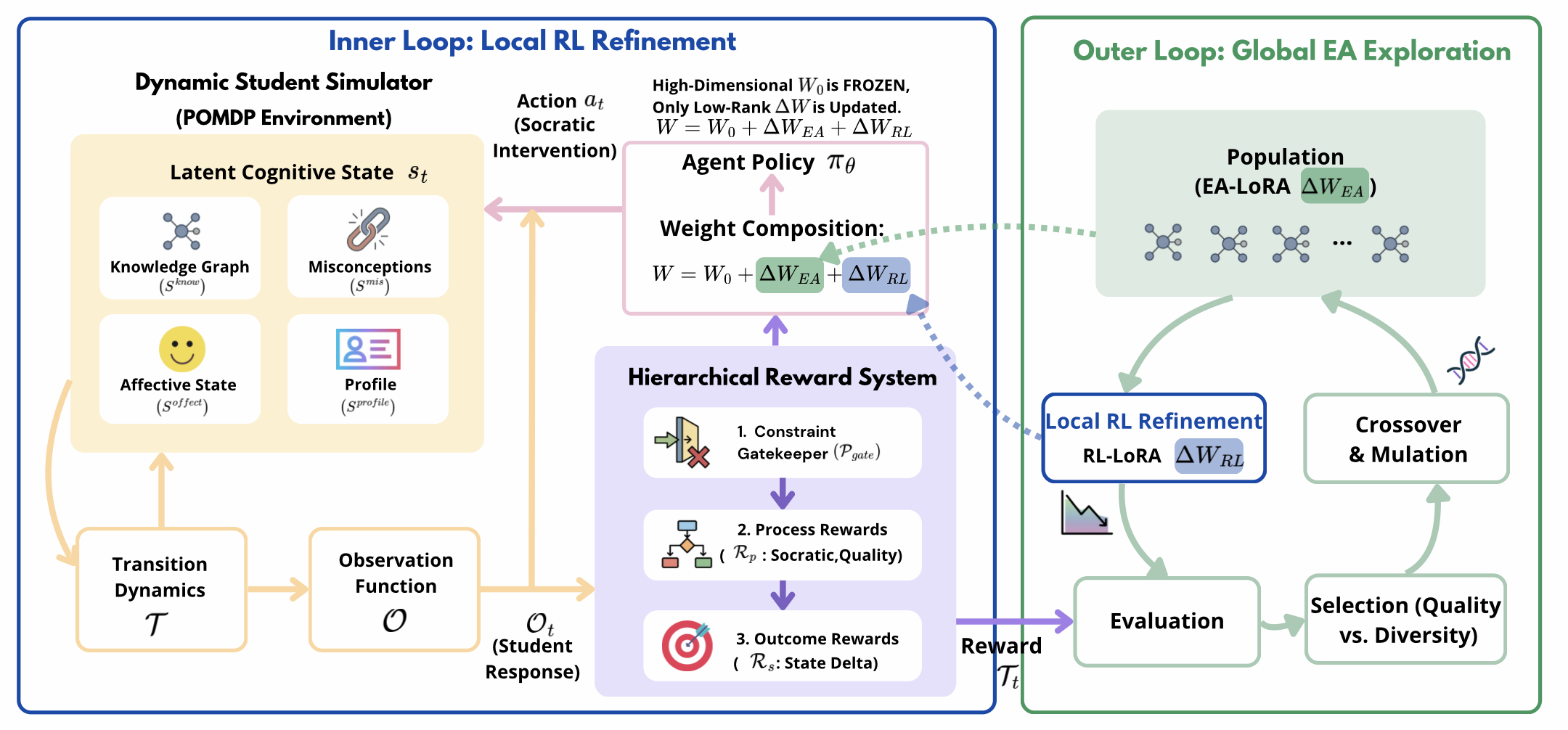}
    \caption{\textbf{The ERL4SIIP Framework.} 
    \textbf{(Left) Dynamic Student Simulator} acts as a surrogate POMDP environment, explicitly modeling latent cognitive states $s_t$ via state-transition dynamics $\mathcal{T}$ to resolve unobservability.
    \textbf{(Middle) Hierarchical   Reward} mitigates sparsity via cascaded evaluation: $\mathcal{P}_{gate}$ enforces safety, 
    $\mathcal{R}_p$ supervises process, and 
   $\mathcal{R}_s$ assesses outcomes.
    \textbf{(Right) LoRA-Division based Optimization} decouples exploration from exploitation $W = W_0 + \Delta W_{EA} + \Delta W_{RL}$. 
    High-Dimensional    $W_0$    is frozen, 
    only Low-Rank   $\Delta W= \Delta W_{EA} + \Delta W_{RL}$    is Updated.
    The outer EA loop evolves global personas $\Delta W_{EA}$ via Pareto selection, while the inner RL loop performs local gradient ascent $\Delta W_{RL}$.}
    \label{fig:framework}
\end{figure*}

To address the optimization challenges of the \Gls{SIIP}, we propose ERL4SIIP. As shown in Figure \ref{fig:framework} and Algorithm \ref{alg:erl4siip}, our framework consists of three coupled components: (1) a Dynamic Student Simulator that serves as the surrogate environment; (2) a Hierarchical Reward System; and (3) a LoRA-Division Based Evolutionary Strategy. In this section, we detail the design of the simulator.


\begin{algorithm}[t]
\caption{LoRA-Division Based ERL Training Procedure}
\label{alg:erl4siip}
\begin{algorithmic}[1]
\REQUIRE Student Simulator $\mathcal{E}$, Base LLM $W_0$, Population Size $N$, Generations $G$, RL Steps $T$.
\REQUIRE Hyperparameters: Ranks $r_{EA}, r_{RL}$, Mutation rate $\sigma$, Reward weights $\lambda$.

\STATE \textbf{Initialize:} Population $\mathcal{P} = \{ \Delta W_{EA}^1, \dots, \Delta W_{EA}^N \}$ randomly.
\STATE \textbf{Initialize:} Shared Replay Buffer $\mathcal{D}$.

\FOR{Generation $g = 1$ to $G$}
    \STATE \textit{// Phase 1: Parallel Local Refinement (Inner RL Loop)}
    \FOR{Individual $i = 1$ to $N$ \textbf{in parallel}}
        \STATE Reset RL-LoRA: $\Delta W_{RL}^i \leftarrow \mathbf{0}$.
        \STATE Construct Agent: $\pi_i \leftarrow W_0 + \Delta W_{EA}^i + \Delta W_{RL}^i$.
        
        \FOR{Step $t = 1$ to $T$}
            \STATE Observe state $b_t$ (Belief) from Simulator $\mathcal{E}$.
            \STATE Sample action $a_t \sim \pi_i(a_t|b_t)$.
            \STATE Receive reward $r_t$ (Eq. 7) and transition to $b_{t+1}$.
            \STATE Store tuple $(b_t, a_t, r_t, b_{t+1})$ in $\mathcal{D}_i$.
            
            \IF{$t \mod K_{update} == 0$}
                \STATE PCGrad Projection: $g_{final} \leftarrow \text{Project}(\{g_k\})$ (Eq. 10).
                \STATE Update RL-LoRA: $\Delta W_{RL}^i \leftarrow \text{Optimizer}(g_{final})$.
            \ENDIF
        \ENDFOR
        \STATE Evaluate Fitness: $F_i \leftarrow \text{AvgReturn}(\pi_i)$.
        \STATE Evaluate Diversity: $D_i \leftarrow \text{Distance}(\Delta W_{EA}^i, \mathcal{P})$.
    \ENDFOR
    
    \STATE \textit{// Phase 2: Global Evolutionary Search (Outer EA Loop)}
    \STATE Select Elites $\mathcal{P}_{elite} \subset \mathcal{P}$ (Eq. 11).
    \STATE $\mathcal{P}_{new} \leftarrow \emptyset$.
    
    \WHILE{$|\mathcal{P}_{new}| < N$}
        \STATE Sample parents $A_{p1}, A_{p2}$ from $\mathcal{P}_{elite}$.
        \STATE Crossover: $A_{child} \leftarrow \gamma A_{p1} + (1-\gamma) A_{p2}$.
        \STATE Mutation: $A_{child} \leftarrow A_{child} + \mathcal{N}(0, \sigma^2 I)$.
        \STATE $\mathcal{P}_{new} \leftarrow \mathcal{P}_{new} \cup \{ A_{child} \}$.
    \ENDWHILE
    \STATE $\mathcal{P} \leftarrow \mathcal{P}_{new}$.
\ENDFOR
\RETURN Best Individual $\pi^*$
\end{algorithmic}
\end{algorithm}

\subsection{Dynamic Student Simulator}
\label{sec:simulator}

\subsubsection{Motivation \& Design Philosophy}
Given the stochastic and unobservable nature of cognitive transitions $\mathcal{T}(s'|s,a)$ defined in Sec. \ref{sec:problem_def}, direct policy optimization on real students is precluded by data scarcity and ethical constraints~\cite{singla2021reinforcement}. Consequently, a high-fidelity surrogate environment is a prerequisite for safe \Gls{RL} deployment. We engineer our simulator to transcend static dialogue generation. Rather than merely predicting next tokens, it maintains a Stateful Cognitive Architecture that explicitly tracks the causal interplay within the latent state $s_t = \langle s^{profile}, s^{know}, s^{mis}, s^{affect} \rangle$. This granular modeling captures distinct learning mechanisms—including memory decay, misconception activation, and affective loops—thereby establishing the rigorous feedback loops essential for optimizing long-horizon Socratic strategies.

\subsubsection{Profile Initialization: Modeling Cognitive Diversity}
To mitigate policy overfitting and prevent mode collapse, we initialize each training episode with a heterogeneous population of student profiles, $s^{profile}$, synthesized via \Glspl{LLM}. This vector functions as a global hyperparameter set governing the simulation dynamics, comprising two distinct categories: (1) \textit{Cognitive Parameters}, specifically Learning Rate ($\alpha \in [0.1, 0.9]$) and \textit{Zone of Proximal Development} width ($w_{zpd}$), which dictate the velocity of knowledge acquisition ($s^{know}$); and (2) \textit{Personality Traits} grounded in the Big Five theory~\cite{goldberg1990alternative}---namely \textit{Curiosity}, \textit{Confidence}, and \textit{Expressiveness}---which modulate proactive interaction patterns and have been historically correlated with academic outcomes~\cite{poropat2009meta}.

\subsubsection{Knowledge \& Misconception Dynamics}
The evolution of the student's cognitive state is modeled as a dynamic interaction between the \textit{Knowledge Graph State} $s^{know}_t$ and the \textit{Misconception State} $s^{mis}_t$. We implement a mechanism where \Glspl{LLM} perform semantic estimation while differential equations govern the state transitions.
\paragraph{Knowledge Mastery Update via LLM-Evaluated ZPD}
For each concept node $c_i$ in the domain knowledge graph, we maintain a mastery level $m_i^{(t)} \in [0, 1]$, which constitutes the vector $s^{know}_t$. The update logic hinges on determining whether the tutor's action $a_t$ falls within the student's \Gls{ZPD}, a core concept in Vygotskian constructivism \cite{vygotsky1978mind}.
Since the "difficulty" of a natural language response is non-trivial to quantify, we employ a frozen LLM as a Cognitive Evaluator.
First, the Evaluator estimates the difficulty scalar $d_{action} \in [0, 1]$ of the tutor's explanation regarding concept $c_i$. Then, the mastery state updates according to a generalized \Gls{IRT} model \cite{lord1980applications}, strictly modulated by the static profile parameters in $s^{profile}$:

\begin{equation}
    m_i^{(t+1)} = m_i^{(t)} + \underbrace{\alpha}_{\in s^{profile}} \cdot \mathbb{I}_{\text{ZPD}}(d_{action}) \cdot (1 - m_i^{(t)})
\end{equation}

Here, $\alpha$ is the student's learning rate retrieved from $s^{profile}$. The indicator function $\mathbb{I}_{\text{ZPD}}$ determines the instructional validity:
\begin{equation}
    \mathbb{I}_{\text{ZPD}}(d_{action}) = 
    \begin{cases} 
    1 & \text{if } m_i^{(t)} \le d_{action} \le m_i^{(t)} + \underbrace{w_{zpd}}_{\in s^{profile}} \\
    0 & \text{otherwise}
    \end{cases}
\end{equation}

By linking $w_{zpd}$ (ZPD width from $s^{profile}$) to the update condition, we mathematically enforce that instructions which are too simple ($d < m$) or too complex ($d > m + w_{zpd}$) result in zero knowledge gain.

\paragraph{The Forgetting Mechanism.}
To simulate the non-monotonic nature of human memory, we implement a passive decay mechanism that operates on $s^{know}_t$ at every time step. Based on the Ebbinghaus Forgetting Curve \cite{ebbinghaus1913memory}, if concept $c_i$ is not activated for $\Delta t$ steps, its mastery decays:
\begin{equation}
    m_i^{(t+\Delta t)} = m_i^{(t)} \cdot e^{-\frac{\Delta t}{S}}
\end{equation}
Crucially, $S$ is the \textit{Memory Stability} parameter derived from the student's static profile $s^{profile}$. This mechanism forces the RL agent to learn \textit{Spaced Repetition} strategies \cite{settles2016trainable}, as neglecting a concept for too long leads to state degradation.

\paragraph{Misconception Activation ("Shadow Graph").}
We explicitly model misconceptions as a discrete binary vector $s^{mis}_t$, where each element corresponds to a specific "bug rule" (e.g., "multiplication always increases numbers") linked to a concept node.
The transition of $s^{mis}_t$ is event-driven, inspired by the theory of "Repair" in procedural skills \cite{vanlehn1990mind}. We utilize the LLM Evaluator to detect pedagogical flaws in the tutor's action $a_t$ (e.g., using ambiguous examples or confirming false premises). If a flaw is detected, the corresponding misconception activates probabilistically:
\begin{equation}
    P(s^{mis}_{k, t+1} = 1 | s^{mis}_{k, t} = 0) = \sigma(\beta \cdot \text{Ambiguity}(a_t) - \gamma \cdot m_k^{(t)})
\end{equation}
where $\text{Ambiguity}(a_t)$ is a score from the LLM Evaluator, and $\gamma$ represents the resilience provided by correct knowledge mastery. This "Shadow Graph" structure allows the simulator to dynamically mimic the competitive interference between correct concepts and stubborn misconceptions.

\subsubsection{Affective State \& Proactive Behavior}
The simulator tracks a 4-dimensional affective vector $s^{affect}_t = \langle \mathcal{F}, \mathcal{E}, \mathcal{L}, \mathcal{B} \rangle$ representing Frustration, Engagement, Cognitive Load, and Boredom.
\paragraph{Affective Transitions.}
The emotional dynamics follow the \textit{Flow Theory} \cite{csikszentmihalyi1990flow}. For instance, Frustration ($\mathcal{F}$) increases when the difficulty of the tutor's question exceeds the student's current mastery plus $w_{zpd}$, while Boredom ($\mathcal{B}$) rises when the content is redundant (i.e., mastery $m_i \to 1$). Dynamics of such affective transitions in tutoring have been widely validated \cite{dmello2012dynamics}.

\paragraph{Proactive Generative Mechanism.}
Crucially, our simulator simulates \textit{student agency}. The observation output $o_t$ (the student's textual response) is generated by an LLM conditioned on the current state tuple. We implement a specific Action Trigger based on intrinsic motivation models \cite{oudeyer2007intrinsic}:
\begin{equation}
    P(\text{ask\_question}) \propto \text{Curiosity} \times \mathcal{F} \times (1 - \text{Confidence})
\end{equation}
This mechanism ensures that the simulated student actively interrupts, expresses confusion, or challenges the tutor, thereby creating the complex, non-linear dialogue trajectories required to train robust Socratic agents.

\subsection{Hierarchical Reward Mechanism}
\label{sec:reward_system}

To mitigate the sparsity of cognitive signals and ensure pedagogical safety, we construct a Hierarchical Reward Mechanism. Unlike naive scalarization, our system enforces strict boundary conditions before optimizing for quality. As defined in Eq. \ref{eq:total_reward}, the total reward $r_t$ is computed via a cascaded evaluation: a "Gatekeeper" first verifies validity, pruning the action space before process and outcome rewards are calculated.

\begin{equation}
    r_t = 
    \begin{cases} 
    - \lambda_c \cdot \mathcal{P}_{gate} & \text{if } \mathcal{P}_{gate} > \tau \quad \text{(Constraint Violation)} \\
    \lambda_p \cdot \mathcal{R}_{p} + \lambda_s \cdot \mathcal{R}_{s} & \text{otherwise} \quad \text{(Effective Intervention)}
    \end{cases}
    \label{eq:total_reward}
\end{equation}

Here, $\mathcal{P}_{gate} \in [0, 1]$ denotes the penalty score, and $\tau \approx 0$ is the tolerance threshold. Valid actions receive a weighted sum of Process Quality ($\mathcal{R}_{p}$) and Student Outcome ($\mathcal{R}_{s}$).

\subsubsection{Layer 1: The Constraint Gatekeeper ($\mathcal{P}_{gate}$)}
This layer enforces the pedagogical "guardrails." We employ an LLM-based verification to quantify penalties as $\mathcal{P}_{gate} = \max(p_{direct}, p_{hallu}, p_{viol})$. The penalty logic targets three specific failures: (1) Knowledge Transmission ($p_{direct}$), which penalizes direct answer provision to enforce the Socratic principle of "guiding, not telling"; (2) Hallucination ($p_{hallu}$), which detects spurious cross-disciplinary links unsupported by the domain Knowledge Graph; and (3) Ethical Violation ($p_{viol}$), ensuring factual accuracy and respectful interaction.

\subsubsection{Layer 2: Pedagogical Process Rewards ($\mathcal{R}_{p}$)}
Upon passing the gatekeeper, we evaluate the \textit{instructional quality} via dense process supervision: $\mathcal{R}_{p} = w_1 \mathcal{R}_{soc} + w_2 \mathcal{R}_{int} + w_3 \mathcal{R}_{pers}$. 
This composite metric integrates: (1) \textit{Socratic Quality} ($\mathcal{R}_{soc}$), where an LLM Critic assesses Cognitive Depth and Recursive Adaptivity following the Paul-Elder taxonomy~\cite{paul2006thinker}; (2) \textit{Interdisciplinary Quality} ($\mathcal{R}_{int}$), rewarding authentic concept linking and real-world task grounding based on Klein's framework~\cite{klein1990interdisciplinarity}, including a trajectory reward for credit assignment; and (3) \textit{Personalization} ($\mathcal{R}_{pers}$), which measures ZPD alignment and affective scaffolding to promote the Flow state.

\subsubsection{Layer 3: Student Outcome Rewards ($\mathcal{R}_{s}$)}
The ultimate objective is measurable student growth. We compute a hybrid outcome signal combining explicit state transitions with semantic analysis:
\begin{equation}
    \mathcal{R}_{s} = \alpha \cdot \mathcal{R}_{state}(\Delta s_t) + \beta \cdot \mathcal{R}_{semantic}(o_t)
\end{equation}
Here, $\mathcal{R}_{state}$ quantifies the explicit reduction in misconception belief mass and mastery gain ($\Delta s^{know}$). Complementarily, $\mathcal{R}_{semantic}$ is derived from an evaluator analyzing the student's response $o_t$ for evidence of high-order competencies: \textit{Transfer}, \textit{Integration}, and \textit{Critical Thinking}.

\subsection{LoRA-Division Based ERL}
\label{sec:erl_framework}

\subsubsection{Motivation: Overcoming Structural Impediments}
Direct optimization of \Glspl{LLM} for Socratic pedagogy faces twin structural barriers. First, the Local Optima Problem: the non-convex landscape of instructional strategies causes gradient-based methods (e.g., PPO) to converge prematurely to risk-averse patterns (e.g., generic encouragement), a phenomenon we term \textit{Strategy Collapse}. Second, the Dimensionality Constraint: applying population-based \Gls{EA} to full-parameter models ($d \approx 10^9$) is computationally intractable and prone to functional degradation under random perturbation. To resolve this, we introduce a LoRA-Division Based architecture that projects the evolutionary search into a low-rank latent manifold, rendering global exploration feasible while preserving the backbone's linguistic capabilities.

\subsubsection{LoRA-Division Based Architecture: Decoupling Exploration \& Exploitation}
We explicitly decompose the weight update $\Delta W$ into two orthogonal low-rank subspaces, physically decoupling global strategy search from local tactical refinement. The composite weight $W$ is formulated as:
\begin{equation}
    W = W_0 + \underbrace{B_{EA} A_{EA}}_{\Delta W_{EA} \text{ (Global/Diversity)}} + \underbrace{B_{RL} A_{RL}}_{\Delta W_{RL} \text{ (Local/Precision)}}
\end{equation}
where $W_0$ is the frozen backbone. The architecture assigns distinct roles:
(1) EA-LoRA ($\Delta W_{EA}$) functions as the population's \textit{Genotype}. These high-rank ($r_{EA}$) adapters encode distinct "Teaching Personas" (e.g., Socratic, Scaffolding) and are updated solely via evolutionary operators to traverse reward basins.
(2) RL-LoRA ($\Delta W_{RL}$) acts as the ephemeral \textit{Phenotype}. Initialized to zero each generation, these low-rank ($r_{RL}$) adapters execute gradient-based "fine-tuning" to maximize local instructional rewards within the subspace defined by the EA persona.

\subsubsection{The ERL Loop}
The training iterates through generations of a population $\mathcal{P} = \{ \pi_1, ..., \pi_N \}$, where each individual is defined by its EA-LoRA weights. The lifecycle comprises three stages:

\paragraph{Step 1: Conflict-Aware Local Optimization (RL-LoRA).}
For each individual, we freeze $\Delta W_{EA}$ and update $\Delta W_{RL}$ via simulator interaction. To resolve the Gradient Conflict among heterogeneous objectives (Safety vs. Exploration), we treat the LoRA parameter space as a multi-objective manifold. Drawing on Conflict-Averse Gradient Descent (PCGrad), we compute individual gradients $g_k = \nabla_{\theta_{RL}} J_k(\pi)$ for each reward component $k$ and apply Gradient Projection before the update:
\begin{equation}
    g_{final} = \sum_{k} \tilde{g}_k, \quad \text{where } \tilde{g}_k = g_k - \sum_{j \neq k} \frac{g_k \cdot g_j}{\|g_j\|^2} \cdot g_j \cdot \mathbb{I}(\cos(g_k, g_j) < 0)
\end{equation}
This projection eliminates destructive interference, ensuring that optimization for pedagogical quality does not compromise safety constraints.

\paragraph{Step 2: Pareto-Based Evolutionary Selection.}
To actively combat mode collapse, we employ a multi-objective selection mechanism. An individual $\pi_i$ is selected as an "Elite" only if it resides on the Local Pareto Front---meaning it is non-dominated across both Performance ($\bar{R}$) and Novelty ($\mathcal{N}$):
\begin{equation}
    \text{Select}(\pi_i) \iff \nexists \pi_j \in \mathcal{P} \text{ s.t. } \bar{R}_j \ge \bar{R}_i \land \mathcal{N}_j \ge \mathcal{N}_i \land (\bar{R}_j > \bar{R}_i \lor \mathcal{N}_j > \mathcal{N}_i)
\end{equation}
This mechanism preserves a diverse "portfolio" of tutors, retaining both safe, steady guides and creative, high-risk explorers.

\paragraph{Step 3: Latent Space Exploration.}
We generate offspring by applying arithmetic operators directly in the EA-LoRA weight space. We employ Arithmetic Crossover ($A_{new} = \gamma A_{p1} + (1-\gamma) A_{p2}$) to perform directional search between high-performing personas, complemented by Gaussian Mutation ($A_{new}' = A_{new} + \mathcal{N}(0, \sigma^2 I)$) to inject stochasticity and escape local optima.

\section{Experiments}
\label{sec:experiments}

In this section, we empirically validate the proposed ERL4SIIP framework. Our experiments aim to answer the following key research questions (RQs):
\begin{itemize}
    \item RQ1 (Simulator Fidelity): Does the student simulator accurately mimic realistic cognitive dynamics and learning behaviors, providing a reliable training environment?
    \item RQ2 (Comprehensive Effectiveness): Does ERL4SIIP outperform baselines in delivering high-quality Socratic instruction that effectively facilitates students' conceptual change and higher-order thinking?
    \item RQ3 (Mechanism \& Robustness): What is the individual contribution of each framework component (e.g., EA loop, Hierarchical Reward), and is the framework efficiently deployable and robust to hyperparameters?
\end{itemize}

\subsection{Experimental Setup}

\subsubsection{Dataset and Benchmarks}
We evaluate our framework on \Gls{SID}~\cite{jiang2025sid}, a benchmark comprising 20,000 annotated multi-turn pedagogical dialogues across diverse STEM domains. We benchmark ERL4SIIP against a comprehensive suite of methods spanning four paradigms: 
(1) General LLMs via Prompting (PE): State-of-the-art foundation models Qwen3~\cite{qwen2025qwen3} and DeepSeek-R1~\cite{deepseekai2025deepseekr1} adapted via few-shot prompting; 
(2) {Education-Specific LLMs: Models tailored for tutoring, including ChatGPT-EDU~\cite{openai_chatgpt_edu}, Gemini-Guiding~\cite{google_gemini_education}, EDUChat~\cite{dan2023educhat}, and InnoSpark\footnote{\url{https://huggingface.co/sii-research/InnoSpark-72B-0710}}; 
(3) SFT Baselines: SocraticLM~\cite{liu2024socraticlm} and SocraticLLM~\cite{ding2024boosting}, representing standard imitation learning; and 
(4) RL Baselines: EduAlign~\cite{song2025cultivating} and Pxplore~\cite{wang2025pxplore}, representing established educational RL approaches without latent state modeling. 
Implementation details are provided in Appendix \ref{app:baselines}.

\subsubsection{Implementation Details}
\label{sec:implementation_details}
We use Qwen3-8B as the backbone model, trained on high-performance GPUs using an efficient parallel evaluation strategy. For our LoRA-Division Based architecture, we set the high-rank EA-LoRA to $r_{EA}=24$ to capture diverse personas, and the low-rank RL-LoRA to $r_{RL}=8$ for local refinement. The evolutionary process maintains a population size of $N=8$ running for $G=20$ generations. The coefficients for our hierarchical reward function are set to $\lambda_c=5.0$ (safety), $\lambda_p=1.0$ (process), and $\lambda_s=2.0$ (outcome).

Further details regarding hardware infrastructure and detailed hyperparameter settings can be found in Appendix \ref{app:implementation}.

\subsection{Validation of the Dynamic Student Simulator (RQ1)}
Before evaluating tutoring agents, establishing the fidelity of our surrogate environment is paramount. We conduct validation from both black-box behavioral and white-box pedagogical perspectives.

\subsubsection{Black-Box Behavioral Analysis (Modified Turing Test)}
To assess the realism of our simulator, we conducted a blinded "Turing Test"~\cite{turing1950computing} against real-world student data.

\paragraph{Ground Truth \& Protocol.} 
We curated a ground-truth dataset from \textit{Dayi}~\cite{dayi2025platform}, a prominent K-12 Q\&A platform, filtering for multi-turn STEM inquiry threads. All data underwent strict anonymization and format normalization to remove platform artifacts (e.g., emojis), ensuring unbiased review. 
We recruited 4 educational experts to evaluate 100 randomized interaction triples. Each triple contained a response to the same tutor input generated by: (1) a Real Student (Dayi), (2) our Dynamic Student Simulator, and (3) a SocraticTeach baseline~\cite{liu2024socraticlm}. Experts were tasked with blindly selecting the response that most authentically reflected K-12 cognitive behaviors.

\begin{table}[h]
    \centering
    \small
    \caption{\textbf{Human Evaluation Results (Turing Test).} Experts select our simulator at a rate statistically comparable to real students, significantly outperforming baseline \cite{liu2024socraticlm}.}
    \label{tab:sim_turing_test}
    \begin{tabular}{lc}
        \hline
        \textbf{Response Source} & \textbf{Expert Selection Rate (\%)} \\ \hline
        Real Student (Ground Truth) & 42.1 \\
        \textbf{Our Simulator (Ours)} & \textbf{41.7} \\
        SocraticTeach Baseline & 17.2 \\ \hline
    \end{tabular}
\end{table}

\paragraph{Results Analysis.}
As detailed in Table \ref{tab:sim_turing_test}, our simulator achieved a selection rate of 41.7\%, which is statistically indistinguishable from real students (42.1\%). This negligible gap confirms that our architecture successfully captures the nuances of student behavior—such as hesitation and partial understanding—making it effectively indistinguishable from reality in a blind setting. In contrast, the baseline (17.2\%) was frequently rejected by experts as appearing overly generic or artificially structured.

\subsubsection{Validation of Automated Evaluation Pipeline}
To legitimize the use of ``LLM-as-a-Judge'' for scalable analysis, we assessed its alignment with human ground truth. We randomly sampled 40 long-horizon trajectories (240 dialogue turns) and tasked 5 educational experts with evaluating them across \textit{Persona Fit}, \textit{Consistency}, and \textit{Learning Dynamics} using a 5-point rubric (Appendix C). 

\paragraph{Alignment Analysis.}
Parallel scoring by GPT-4-Turbo demonstrates robust fidelity (Table \ref{tab:human_llm_alignment}). The automated evaluator achieves a Pearson correlation of $r=0.88$ ($p < 0.001$) with human scores, significantly exceeding the reliability threshold ($r > 0.8$) established in recent benchmarking studies~\cite{liu2023geval}. Furthermore, a Cohen's Kappa of $\kappa=0.75$ indicates ``Substantial Agreement,'' mirroring the consensus levels typically observed among human experts~\cite{zheng2023judging}. With a negligible average score divergence ($\Delta \le 0.3$), these metrics validate the judge's reliability for subsequent large-scale experiments.

\begin{table}[h]
    \centering
    \small
    \caption{\textbf{Human-LLM Alignment Check.} Average scores (1-5 scale) on 40 trajectories show high consistency. The LLM judge achieves a Pearson's $r=0.88^{**}$ and Cohen's $\kappa=0.75$, indicating substantial agreement with experts.}
    \label{tab:human_llm_alignment}
    \begin{tabular}{l|ccc}
        \hline
        \textbf{Evaluator} & Persona & Consistency & Dynamics  \\ \hline
        Human Experts & 3.7 & 3.8 & 3.6  \\
        LLM Judge & 3.8 & 4.1 & 3.7  \\ 
        \textit{Difference} & \textit{+0.1} & \textit{+0.3} & \textit{+0.1}  \\ \hline
    \end{tabular}
\end{table}

\subsection{Main Results (RQ2)}
\label{sec:main_results}
We evaluate ERL4SIIP via a dual-lens protocol: Instructional Quality (Table \ref{tab:main_results}) tracks teacher-side strategic fidelity, while Student Learning Outcomes (Table \ref{tab:hope_scores}) quantify the resultant cognitive shifts. To rigorously validate deep conceptual change, we formulate the \Gls{HOPE} metric suite, grounded in constructivist theory: 
(1) Knowledge Integration (KI) adopts Linn's framework~\cite{linn2006knowledge} to measure the coherence of synthesized connections between fragmented ideas; 
(2) Knowledge Transfer (KT) operationalizes Perkins' ``High-road Transfer''~\cite{perkins1988teaching} to assess the abstraction of principles for novel cross-disciplinary contexts; 
(3) Critical Thinking (CT) targets ``Self-Regulation'' and ``Inference'' behaviors defined in the Delphi Report~\cite{facione1990critical}; and 
(4) Creativity (CR) evaluates ``mini-c'' creativity~\cite{beghetto2007toward}, focusing on the student's generation of personally novel conceptual links.

\subsubsection{Comparative Performance Analysis}
As detailed in Tables \ref{tab:main_results} and \ref{tab:hope_scores}, ERL4SIIP consistently outperforms baselines across both instructional quality and student outcomes. We analyze performance across model paradigms:

\paragraph{SFT-Only: The Imitation Trap.}
Models relying solely on behavior cloning (e.g., SocraticLM) capture surface-level dialogue patterns but falter in strategic reasoning. Their low \textit{Level-3 Guidance Rate} (33.28\%) and \textit{Blueprint Planning} scores (17.69) reveal a tendency to passively respond rather than actively guide. Consequently, students interacting with SFT agents exhibit the lowest Critical Thinking counts (1.22), confirming that imitation fails to induce the "productive struggle" necessary for high-order cognition.

\paragraph{RL-Only \& General LLMs: The Pedagogy Deficit.}
Standard RL (e.g., Pxplore) and general foundation models (e.g., DeepSeek-V3) improve upon SFT but lack targeted pedagogical alignment. While maintaining better conversational flow, they struggle to resolve misconceptions (lower 3C scores), resulting in mediocre Knowledge Integration rates ($\approx 40\%$). Without latent state tracking, these models fail to deliver the precise interventions required for deep conceptual change.

\paragraph{Education-Specific LLMs: Scale vs. Strategy.}
Commercial models like ChatGPT-EDU set a strong baseline, leveraging massive scale to achieve respectable Knowledge Integration ($>50\%$). However, they lag significantly in \textit{Strategy Diversity} (31.97 vs. 36.80 for Ours) and \textit{Socratic Consistency}. This suggests that while scale enables broad knowledge retrieval, it cannot replace the specific, structured scaffolding provided by ERL4SIIP, as evidenced by the lower student Critical Thinking scores (2.85).

\subsubsection{Mechanism Analysis}
The empirical superiority of ERL4SIIP stems from its structural innovations, directly mapping architectural features to student cognitive growth:

\noindent (1) Diversity Drives Critical Thinking. 
A critical advantage of our evolutionary approach is high \textit{Strategy Diversity} (SV 36.80). By maintaining a diverse population of teaching personas, ERL4SIIP prevents mode collapse, forcing simulated students to constantly re-evaluate premises. This adaptivity yields the highest \textit{Critical Thinking} frequency (3.85)---nearly triple the rate of SFT baselines.

\noindent (2) Precision Scaffolding Drives Transfer. 
The hierarchical reward ensures interdisciplinary links are logically sound (Highest X-SRG 4.92). This precise scaffolding (L3 GR 82.50\%) effectively guides students to map concepts across domains, culminating in superior \textit{Knowledge Transfer} (KT 4.45).

\noindent (3) Long-term Planning Drives Creativity. 
High \textit{Socratic Consistency} (65.40\%) reflects the agent's ability to maintain a coherent pedagogical plan. This sustained guidance provides the "safe structure" students need to explore novel hypotheses, reflected in superior \textit{Creativity} scores (CR 4.35). Notably, the w/o PRM ablation (SC 30.20\%) confirms that without dense process supervision, agents revert to myopic optimization, losing the capacity for Socratic inquiry.

\begin{table*}[t!]
\centering
\small
\caption{\textbf{Instructional Quality (Teacher-Side).} ERL4SIIP achieves state-of-the-art performance across 5 subjective (1-5 scale) and 7 objective metrics, particularly in Socratic guidance (L3 GR) and Strategy Diversity (SV).}
\label{tab:main_results}
\resizebox{0.95\textwidth}{!}{%
\begin{tabular}{llcccccccccccc}
\toprule
\multirow{2}{*}{\textbf{Category}} & \multirow{2}{*}{\textbf{Algorithm}} & \multicolumn{5}{c}{\textbf{Subjective Indicators (Auto, 1-5 $\uparrow$)}} & \multicolumn{7}{c}{\textbf{Objective Indicators (Auto, $\uparrow$)}} \\
\cmidrule(lr){3-7} \cmidrule(lr){8-14}
 & & \textbf{X-SRG} & \textbf{M-RCC} & \textbf{X-MSR} & \textbf{IRA} & \textbf{TCF} & \textbf{SD (\%)} & \textbf{SV} & \textbf{IKT} & \textbf{BP} & \textbf{L3 GR (\%)} & \textbf{SC (\%)} & \textbf{3C} \\
\midrule
\multirow{2}{*}{SFT-Only} & SocraticLM & 2.11 & 1.75 & 1.62 & 1.77 & 1.69 & 97.32 & 25.47 & 17.95 & 17.69 & 33.28 & 32.19 & 48.77 \\
 & SocraticLLM & 3.32 & 2.97 & 2.05 & 2.11 & 1.75 & 97.51 & 24.93 & 18.77 & 21.29 & 39.25 & 27.58 & 45.26 \\
\midrule
\multirow{2}{*}{RL-Only} & EduAlign & 4.09 & 4.01 & 4.17 & 3.93 & 4.21 & 98.23 & 29.75 & 19.77 & 23.28 & 65.49 & 42.88 & 55.26 \\
 & Pxplore & 4.03 & 3.92 & 4.09 & 3.81 & 4.17 & 98.54 & 27.33 & 17.56 & 24.87 & 64.51 & 45.29 & 69.73 \\
\midrule
\multirow{4}{*}{Edu-LLMs} & InnoSpark & 4.74 & 4.11 & 4.23 & 4.46 & 4.39 & 99.15 & 31.40 & 19.81 & 21.34 & 68.52 & 41.27 & 57.29 \\
 & ChatGPT-EDU & 4.89 & 4.35 & 4.47 & 4.63 & 4.52 & 99.23 & 31.97 & 47.59 & 35.26 & 79.73 & 58.18 & 73.62 \\
 & Gemini-Guiding & 4.85 & 4.43 & 4.52 & 4.47 & 4.59 & 98.89 & 35.44 & 45.28 & 38.27 & 78.85 & 62.13 & 75.81 \\
 & EduChat & 4.32 & 3.57 & 3.22 & 3.74 & 4.09 & 95.41 & 29.81 & 11.53 & 19.33 & 66.34 & 34.57 & 55.32 \\
\midrule
\multirow{2}{*}{General LLMs} & Qwen3 & 4.03 & 3.81 & 3.79 & 3.88 & 3.92 & 97.63 & 31.27 & 17.56 & 23.20 & 44.33 & 34.58 & 62.74 \\
 & DeepSeek-V3 & 4.17 & 4.05 & 3.81 & 3.89 & 4.01 & 98.27 & 30.93 & 25.48 & 27.29 & 51.83 & 44.32 & 65.41 \\
\midrule
\midrule
\multirow{3}{*}{\textbf{Ours}} & ERL (w/o LoRA-Division) & 4.79 & 4.25 & 4.35 & 4.55 & 4.45 & 99.10 & 33.50 & 46.12 & 34.80 & 75.20 & 55.60 & 73.10 \\
 & ERL (w/o PRM) & 3.55 & 3.10 & 2.85 & 3.05 & 3.20 & 96.50 & 26.10 & 22.30 & 19.50 & 40.50 & 30.20 & 50.10 \\
 & \textbf{ERL4SIIP (Full)} & \textbf{4.92} & \textbf{4.51} & \textbf{4.60} & \textbf{4.71} & \textbf{4.65} & \textbf{99.50} & \textbf{36.80} & \textbf{49.50} & \textbf{40.10} & \textbf{82.50} & \textbf{65.40} & \textbf{78.90} \\
\bottomrule
\end{tabular}%
}
\end{table*}

\begin{table}[t!]
\centering
\small
\caption{\textbf{Student Learning Outcomes (HOPE Scores).} Impact of tutoring agents on student cognitive growth in the simulator. ERL4SIIP achieves significantly higher CT and KT.}
\label{tab:hope_scores}
\resizebox{\columnwidth}{!}{%
\begin{tabular}{llcccc}
\toprule
\multirow{2}{*}{\textbf{Category}} & \multirow{2}{*}{\textbf{Algorithm}} & \multicolumn{4}{c}{\textbf{Higher-Order Indicators (HOPE, $\uparrow$)}} \\
\cmidrule(lr){3-6}
 & & \textbf{KI (\%)} & \textbf{KT (1-5)} & \textbf{CT (Count)} & \textbf{CR (1-5)} \\
\midrule
SFT-Only & SocraticLM & 35.21 & 2.85 & 1.22 & 2.54 \\
\midrule
RL-Only & EduAlign & 42.88 & 3.51 & 2.15 & 3.12 \\
\midrule
\multirow{2}{*}{Edu-LLMs} & ChatGPT-EDU & 52.15 & 4.12 & 2.85 & 4.21 \\
 & Gemini-Guiding & 51.33 & 4.05 & 2.78 & 4.15 \\
\midrule
General LLMs & DeepSeek-V3 & 43.55 & 3.58 & 2.35 & 3.68 \\
\midrule
\midrule
\multirow{3}{*}{\textbf{Ours}} & ERL (w/o LoRA-Division) & 54.20 & 4.28 & 3.22 & 4.08 \\
 & ERL (w/o PRM) & 38.50 & 3.05 & 1.55 & 2.85 \\
 & \textbf{ERL4SIIP (Full)} & \textbf{58.12} & \textbf{4.45} & \textbf{3.85} & \textbf{4.35} \\
\bottomrule
\end{tabular}%
}
\end{table}
\subsection{Ablation Study (RQ3)}
\label{sec:mechanism_analysis}

Beyond performance metrics, we dissect the mechanistic underpinnings of ERL4SIIP. We validate our core design philosophy—the synergy of hierarchical supervision with global exploration, and the structural decoupling of strategy and tactic—through a triangulation of ablation studies, theoretical rank sensitivity analysis, and latent space visualization.

\subsubsection{Deconstructing Component Contributions}
The ablation results (referencing Table \ref{tab:main_results}) isolate the distinct roles of the hierarchical reward and the Evolutionary loop.

\paragraph{PRM as the Pedagogical Compass.}
The comparison with \textit{w/o PRM} exposes the fundamental inadequacy of sparse outcome-based RL for long-horizon tutoring. Without dense supervision, the agent suffers from the "credit assignment problem," unable to attribute final success to specific turn-level actions. This leads to a catastrophic drop in \textit{Socratic Consistency} (65.40\% $\to$ 30.20\%), confirming that the PRM acts as a necessary compass to guide the agent through the vast action space toward coherent scaffolding structures.

\paragraph{EA as the Diversity Engine.}
Standard RL baselines (e.g., Pxplore) succumb to mode collapse, converging to local optima with low \textit{Strategy Diversity} ($\sim$27\%). In contrast, the evolutionary loop serves as a continuous "escape mechanism." By maintaining a heterogeneous population, EA explores strategic regions far from the initialization manifold. This is quantitatively confirmed by ERL4SIIP's superior Strategy Diversity (36.80) and Creativity (4.35), demonstrating that global search is indispensable for discovering the adaptive repertoire that single-point optimization misses.

\subsubsection{Theoretical Insight: The Rank-Task Matching Principle}
\label{sec:synergy}
A central challenge in tutoring agents is the "Comprehensive Paradox": optimizing for broad adaptivity often leads to "Average Mode Collapse". Our LoRA-Division Based architecture resolves this via an information-theoretic allocation of parameter budget ($r_{EA} \gg r_{RL}$), as analyzed in Table \ref{tab:rank_sensitivity}.

Our sensitivity analysis reveals a governing principle for hierarchical alignment:
(1) High-Rank Exploration ($r_{EA}=24$): Constructing distinct "Teaching Personas" (e.g., \textit{Provocateur} vs. \textit{Scaffolder}) requires global semantic reconstruction, a task with high \textit{intrinsic dimensionality}. High-rank adapters provide the necessary capacity to encode these non-local stylistic features. Reducing this capacity ($r_{EA}=4$) leads to "Strategic Myopia," trapping the agent in its initial distribution.
(2) Low-Rank Exploitation ($r_{RL}=8$): Conversely, tactical refinement lies on a lower-dimensional manifold. A lower rank acts as a \textit{regularization constraint}, forcing PPO updates to operate strictly within the subspace defined by the EA persona, ensuring precise execution without overwriting the fundamental strategic stance.
The 24/8 split empirically strikes the optimal balance, securing near-peak diversity alongside highest guidance depth.

\begin{table}[h]
\centering
\normalsize
\caption{\textbf{Sensitivity Analysis of Rank Allocation (Total Rank = 32).} Comparing different $r_{EA}/r_{RL}$ splits reveals that the "EA-Dominant" configuration (24/8) achieves the optimal balance, maximizing both teacher strategy diversity and comprehensive student outcomes.}
\label{tab:rank_sensitivity}
\resizebox{\columnwidth}{!}{%
\begin{tabular}{l|cc|cccc}
\toprule
\textbf{Configuration} & \multicolumn{2}{c|}{\textbf{Teacher Action Metrics}} & \multicolumn{4}{c}{\textbf{Student HOPE Metrics (Outcomes)}} \\
($r_{EA}$ / $r_{RL}$) & \textbf{SV (\%)} & \textbf{L3 GR (\%)} & \textbf{KI (\%)} & \textbf{KT (1-5)} & \textbf{CT (Count)} & \textbf{CR (1-5)}  \\
\midrule
Single LoRA & 33.50 & 75.20 & 54.20 & 4.28 & 3.22 & 4.08  \\
\midrule
\textit{EA-Dominant} & & & & & &  \\
28 / 4 & \textbf{37.15} & 81.50 & \textbf{58.50} & 4.38 & 3.75 & \textbf{4.45}  \\
\textbf{24 / 8 (Ours)} & 36.80 & \textbf{82.50} & 58.12 & \textbf{4.45} & \textbf{3.85} & 4.42  \\
\midrule
\textit{Balanced} & & & & & &  \\
16 / 16 & 35.20 & 79.80 & 56.50 & 4.32 & 3.65 & 4.25  \\
\midrule
\textit{RL-Dominant} & & & & & &  \\
8 / 24 & 31.50 & 78.50 & 52.10 & 4.15 & 3.45 & 3.95  \\
4 / 28 & 28.40 & 76.20 & 48.50 & 3.95 & 3.10 & 3.70  \\
\bottomrule
\end{tabular}%
}
\end{table}

\subsubsection{Visual Validation}
To visually corroborate these findings, we projected the effective LoRA weights ($\Delta W_{EA} + \Delta W_{RL}$) of elite individuals into a 2D landscape using t-SNE (Figure \ref{fig:latent_space_trajectory}).

\begin{figure}[t]
    \centering
    \includegraphics[width=0.48\textwidth]{pic/placeholder_latent_trajectory-v1.png}
    \caption{\textbf{Optimization Trajectories in the Latent Reward Landscape.} (Yellow = High Reward, Blue = Low). 
    \textbf{(Top-Left) PPO Baseline:} Trapped in local optima due to gradient dependency.
    \textbf{(Top-Right) ERL4SIIP (Ours):} Achieves directional global exploration, jumping to the global peak.
    \textbf{(Bottom-Left) w/o PRM:} Exhibits chaotic, unguided dispersion.
    \textbf{(Bottom-Right) w/o LoRA-Division:} Shows weak directionality.}
    \label{fig:latent_space_trajectory}
\end{figure}

The visualization provides compelling evidence for our architectural claims:
(1) Global vs. Local Search: The PPO Baseline (Top-Left) exhibits typical "Local Entrapment," rapidly stagnating in the nearest rugged basin. In sharp contrast, ERL4SIIP (Top-Right) demonstrates "Directional Global Exploration," where EA enables long-range jumps to escape traps while RL performs efficient local climbing.
(2) Component Necessity: The w/o PRM plot (Bottom-Left) reveals "Unguided Stochastic Dispersion" due to sparse signals, while w/o LoRA-Division Based (Bottom-Right) shows "Coupled Interference," where competing gradients confound the search direction, leading to suboptimal convergence.
\section{Conclusion and Future Work}
\label{sec:conclusion}

In this paper, we propose ERL4SIIP, a novel Evolutionary Reinforcement Learning (ERL) framework specifically tailored for the Socratic Interdisciplinary Instructional Problem (SIIP). 
 ERL4SIIP integrates:
(1) a dynamic student simulator grounded in a STEM knowledge graph for latent state modeling; (2) a Hierarchical Reward Mechanism that decomposes long-horizon goals into dense signals; and
(3) a LoRA-Division based optimization strategy coupling evolutionary algorithms for population-level global search with PPO for local gradient ascent.
Experimental results demonstrate that
ERL4SIIP significantly outperforms state-of-the-art baselines in
improving students’ higher-order abilities—particularly knowledge
integration and interdisciplinary transfer—while exhibiting richer
teaching strategy diversity and superior robustness against the
"spoon-feeding" trap.

Future research will focus on two axes. 
First, We will study the
cooperation of EA and contrastive feedback based
RL methods such as
Direct Preference Optimization (DPO) or Group Relative Policy Optimization (GRPO). 
Second, We plan to apply the framework 
to real-world Project-based learning scenarios.

\bibliographystyle{ACM-Reference-Format}
\bibliography{lib}

\newpage
\appendix
\section{Detailed Problem Formulation}

\subsection{State Space Specifications}
\label{app:state_details}
The latent state vector $s_t$ defined in Section \ref{sec:problem_def} consists of the following detailed components:

\paragraph{Static Profile ($s^{profile}$)} This vector contains invariant parameters that modulate the transition dynamics:
\begin{itemize}
    \item Cognitive Baseline: Includes the \textit{Learning Rate} ($\alpha$) and the width of the \textit{Zone of Proximal Development} ($w_{zpd}$), which determine the magnitude of knowledge updates
    \item Personality Traits: Based on the Big Five theory, we explicitly model \textit{Curiosity}, \textit{Confidence}, and \textit{Expressiveness}. These parameters influence the probability of the student initiating proactive questioning (as detailed in the Student Simulator section).
\end{itemize}

\paragraph{Affective State ($s^{affect}_t$)} A 4-dimensional continuous vector tracking emotional dynamics integral to learning:
\begin{equation}
    s^{affect}_t = \langle \text{Frustration}, \text{Engagement}, \text{Cognitive Load}, \text{Boredom} \rangle
\end{equation}
These dimensions evolve according to Flow Theory , where mismatches between task difficulty and student skill trigger transitions between states (e.g., high difficulty relative to skill increases Frustration).

\subsection{Belief State Update}
\label{app:belief_update}
Given the POMDP structure, the agent updates its belief state $b_t(s)$ at each time step upon observing $o_{t+1}$ after taking action $a_t$.The update follows the standard Bayesian filter:
\begin{equation}
    b_{t+1}(s') \propto \mathcal{O}(o_{t+1} | s', a_t) \sum_{s \in \mathcal{S}} \mathcal{T}(s' | s, a_t) b_t(s)
\end{equation}
where $\mathcal{O}$ is the observation function likelihood and $\mathcal{T}$ is the state transition probability. In our ERL4SIIP framework, this explicit belief tracking allows the agent to make decisions based on the inferred probability of specific misconceptions being active, rather than relying solely on surface-level dialogue history.

\section{Detailed Baseline Descriptions}
\label{app:baselines}

We benchmark ERL4SIIP against a comprehensive suite of methods spanning four distinct paradigms:

\paragraph{1. General LLMs with \Gls{PE} (PE).}
We evaluate foundation models via zero-shot prompting to establish a performance floor without task-specific tuning. Qwen3~\cite{qwen2025qwen3} serves as a state-of-the-art open-source benchmark, while DeepSeek-R1~\cite{deepseekai2025deepseekr1} is included to evaluate the impact of specialized reasoning capabilities on Socratic inquiry.

\paragraph{2. Education-Specific LLMs.}
These models utilize domain-specific corpora to enhance tutoring aptitude. 
ChatGPT-EDU~\cite{openai_chatgpt_edu} and Gemini-Guiding~\cite{google_gemini_education} represent commercial variants explicitly optimized for academic scenarios. On the open-source side, EDUChat~\cite{dan2023educhat} provides a specialized dialogue backbone, and InnoSpark\footnote{\url{https://huggingface.co/sii-research/InnoSpark-72B-0710}} offers instruction-following tuning tailored for domain-specific educational tasks.

\paragraph{3. Supervised Fine-Tuning (SFT-Only).}
Representing the Behavior Cloning paradigm, these models learn directly from expert demonstrations. SocraticLM~\cite{liu2024socraticlm} is fine-tuned on personalized Socratic dialogues to acquire questioning patterns, while SocraticLLM~\cite{ding2024boosting} specifically integrates Socratic methodology for mathematical conversations.
\begin{table}[htbp] 
\centering
\normalsize
\caption{\textbf{Training Efficiency Comparison on Qwen3-8B (8x H100).} We compare full-parameter evolution against standard single LoRA and our LoRA-Division Based approach. "Standard ERL (Single LoRA)" uses a single LoRA module with rank $r=r_{EA}+r_{RL}=32$. Both LoRA methods enable parallel evaluation of 8 offspring.}
\label{tab:efficiency_app1}
\resizebox{\columnwidth}{!}{%
\begin{tabular}{l|c|cc}
\hline
\textbf{Method} & \textbf{Trainable Params} & \textbf{Peak VRAM / GPU} & \textbf{Avg. Time / Generation} \\ \hline
Full-Parameter ERL & $\sim$8.0 B & \textit{OOM (>80GB)} & \textit{N/A} \\
Standard ERL (Single LoRA) & $\sim$160 M & 24.5 GB & 19 h \\
\textbf{LoRA-Division Based ERL (Ours)} & $\sim$160 M & \textbf{21.8 GB} & \textbf{11 h} \\ \hline
\end{tabular}%
}
\end{table}
\paragraph{4. Reinforcement Learning (RL-Only).}
These baselines employ standard RL optimization without latent state modeling. EduAlign~\cite{song2025cultivating} utilizes multi-dimensional reward modeling for pedagogical alignment, and Pxplore~\cite{wang2025pxplore} focuses on strategy exploration for personalized path planning.

\section{Additional Implementation Details}
\label{app:implementation}
\begin{table*}[h]
\centering
\scriptsize
\renewcommand{\arraystretch}{1.3} 
\caption{\textbf{Comparative Case Study: Scaffolding vs. Spoon-feeding.} Comparison of instructional strategies for student Sun Meng (Profile: Rote Learner). While the Baseline (Left) relies on efficiency, ERL4SIIP (Right) prioritizes cognitive struggle. \good{Green text} indicates high-order cognitive activation; \bad{Red text} indicates passive reception.}
\label{tab:case_study_comparison}
\begin{tabular}{p{0.48\linewidth} | p{0.48\linewidth}}
\toprule
\multicolumn{1}{c|}{\textbf{\large Baseline: ChatGPT-EDU}} & \multicolumn{1}{c}{\textbf{\large Ours: ERL4SIIP}} \\
\multicolumn{1}{c|}{\textcolor{badred}{\textit{(The Efficient Tutor)}}} & \multicolumn{1}{c}{\textcolor{goodgreen}{\textit{(The Strategic Guide)}}} \\
\midrule
\multicolumn{2}{p{0.96\linewidth}}{\textbf{Student Profile (Sun Meng):} ``Rote Learner''. Strong declarative memory but holds misconception: \textit{``Ferns differ from flowering plants only by flowers.''}} \\
\midrule

\multicolumn{2}{c}{\cellcolor{black!5}\textbf{Round 1: Initial Inquiry}} \\
\multicolumn{2}{p{0.96\linewidth}}{\cellcolor{black!5}\textit{Student Action:} (Holding a fern) ``Teacher, when does this flower? I want to draw it in the `Flowering Plants' section.''} \\
\textbf{R1 Teacher (EDU):} ``That's a good observation. But this is a kidney fern. Do you remember which type of plant has roots/stems/leaves but reproduces differently?'' & \textbf{R1 Teacher (ERL4SIIP):} ``You noticed its roots and stems. But to determine if it `flowers', we can't just look at the front. \good{Turn the leaf over}, draw what you see, and tell me.'' \\
\textit{\textbf{Analysis:} \neu{Standard Recall Trigger}. Verbal-only interaction.} & \textit{\textbf{Analysis:} \good{Action-Based Scaffolding}. Forces physical observation to break misconception anchor.} \\
\midrule

\textbf{R2 Student:} ``Um... Ferns? But don't they flower?'' & \textbf{R2 Student:} (Turns leaf) ``Eh? The back is full of brown dots... disgusting...'' \\
\textbf{R2 Teacher (EDU):} ``Correct! \bad{They indeed don't flower.} Look at the back, \bad{those are `spores'.} And look, this shady corner is what ferns like.'' & \textbf{R2 Teacher (ERL4SIIP):} ``Those aren't eggs. Activate your memory: What is the brown structure in `Plants without seeds'? Hint: It acts like a seed but \good{needs water}.'' \\
\textit{\textbf{Analysis:} \bad{Direct Spoon-feeding}. The agent explicitly provides both the biological fact (spores) and environmental link.} & \textit{\textbf{Analysis:} \good{Knowledge Activation}. Utilizes rote memory strength to retrieve the concept without telling.} \\
\midrule

\textbf{R3 Student:} ``Oh! So those are spores. I remember now: Fern, no flowers, likes humidity.'' & \textbf{R3 Student:} ``Spores... Needs water! Ah, it's a fern! But does this relate to where it grows?'' \\
\textbf{R3 Teacher (EDU):} ``Great summary! Write it down. Next topic...'' & \textbf{R3 Teacher (ERL4SIIP):} ``That is the key. Since spores `need water', use your \good{Geography Knowledge}. Look at this tap and shade. \good{Why here} and not the playground?'' \\
\textit{\textbf{Analysis:} \bad{Shallow Integration}. Student accepts facts passively. Interaction ends prematurely.} & \textit{\textbf{Analysis:} \good{Cross-Disciplinary Inference}. Forces student to deduce the link: Bio Needs $\leftrightarrow$ Geo Evidence.} \\
\midrule

\multirow{4}{*}{\textit{(Session ended in Round 3)}} & \textbf{R4 Student:} \good{[Aha Moment!]} ``I get it! The playground is dry. Here is humid... so it suits spores! The book definition actually means real life!'' \\
 & \textbf{R4 Teacher (ERL4SIIP):} ``Exactly. You discovered its survival logic.'' \\
 & \textit{\textbf{Analysis:} \good{Deep Knowledge Integration}. Self-generated bridging of disciplines.} \\
\midrule

\multicolumn{1}{c|}{\cellcolor{badred!5}\textbf{Outcome Summary}} & \multicolumn{1}{c}{\cellcolor{goodgreen!5}\textbf{Outcome Summary}} \\
\xmark\ Misconception corrected via \bad{Direct Telling}. & \cmark\ Misconception resolved via \good{Guided Observation}. \\
\xmark\ Integration: \bad{Passive Acceptance}. & \cmark\ Integration: \good{Active Construction}. \\
\neu{(!)} Engagement: Medium (Low Cognitive Load). & \cmark\ Engagement: \good{High (``Aha!'' Moment)}. \\
\bottomrule
\end{tabular}
\end{table*}

\paragraph{Hardware and Parallelization Strategy.}
Our experiments are conducted on a high-performance computational cluster equipped with 8 $\times$ NVIDIA H100 (80GB HBM3) GPUs. To accelerate the evolutionary process, we implement an efficient parallel evaluation strategy: within each evolutionary generation, 8 individual offspring (agents) are deployed concurrently across the GPUs to interact with their respective student simulators and perform local RL updates. This setup significantly reduces the wall-clock time required for population-based training.

\paragraph{Detailed Hyperparameters.}
Beyond the core parameters mentioned in the main text, we provide further settings for reproducibility.
Within the inner RL loop, PPO updates occur every $2048$ interaction steps (covering extended dialogue horizons) with a clipping range of $0.2$. The generalized advantage estimation (GAE) parameters are set typically at $\gamma=0.99$ and $\lambda=0.95$. The learning rate for the RL-LoRA component is set to $1e-5$ with a cosine decay schedule.
\begin{table*}
\centering
\caption{\textbf{The Multi-Dimensional Scoring Rubric for Expert Evaluation.} Experts assessed simulated student trajectories on a 1-5 scale across three dimensions.}
\label{tab:expert_rubric_app}
\resizebox{\textwidth}{!}{%
\begin{tabular}{l|l|l}
\hline
\textbf{Dimension} & \textbf{Core Definition \& Evaluation Focus} & \textbf{Anchor Rating Descriptions (Simplified)} \\ \hline
\multirow{3}{*}{\begin{tabular}[c]{@{}l@{}}\textbf{D1: Persona Fit \&}\\\textbf{Anthropomorphism}\end{tabular}} & \multirow{3}{*}{\begin{tabular}[c]{@{}l@{}}Does the simulator avoid generic "AI flavor" (e.g., overly structured,\\ verbose) and accurately reflect the specific profile's linguistic style,\\ pacing, and emotional state?\end{tabular}} & \textbf{1 (AI-like):} acts like an encyclopedia or chatbot; overly polite/structured. \\
 &  & \textbf{3 (Generic):} plausible student but lacks distinct persona traits. \\
 &  & \textbf{5 (Immersive):} perfectly captures psychological traits and non-verbal cues. \\ \hline
\multirow{3}{*}{\begin{tabular}[c]{@{}l@{}}\textbf{D2: Cognitive \&}\\\textbf{Behavioral Consistency}\end{tabular}} & \multirow{3}{*}{\begin{tabular}[c]{@{}l@{}}Are the student's behaviors logically self-consistent? Are skill\\ boundaries stable, avoiding unexplained "skill splits" (solving hard\\ tasks while failing easy ones) or character drift?\end{tabular}} & \textbf{1 (Broken Logic):} unexplained contradictions in ability or personality. \\
 &  & \textbf{3 (Minor Flaws):} mostly consistent with occasional unnatural fluctuations. \\
 &  & \textbf{5 (Highly Coherent):} knowledge gaps and error patterns form a consistent loop. \\ \hline
\multirow{3}{*}{\begin{tabular}[c]{@{}l@{}}\textbf{D3: Learning Dynamics}\\\textbf{\& Trajectory Plausibility}\end{tabular}} & \multirow{3}{*}{\begin{tabular}[c]{@{}l@{}}Does the learning process align with cognitive theories (e.g., ZPD)?\\ Does it reflect realistic non-linear growth, including struggles,\\ "aha moments", rather than mechanical step-wise jumps?\end{tabular}} & \textbf{1 (Mechanical):} instant perfect answers after a hint; ignores scaffolding. \\
 &  & \textbf{3 (Linear):} steady, predictable progress lacking realistic struggle. \\
 &  & \textbf{5 (Realistic Dynamic):} perfectly models non-linear growth with genuine struggle/insight. \\ \hline
\end{tabular}%
}
\end{table*}
\section{Detailed Expert Evaluation Rubric}
\label{app:rubric}

Table \ref{tab:expert_rubric_app} details the multi-dimensional scoring rubric used by human experts to evaluate the fidelity of simulated student trajectories. The rubric focuses on avoiding generic AI characteristics and ensuring pedagogical plausibility across persona, cognition, and learning dynamics.
\section{Qualitative Case Study Details}
\label{app:case_study}

This appendix provides the full qualitative case study referencing the quantitative findings in Section \ref{sec:main_results}.

\subsection{Case Study}
To visualize the pedagogical drivers of high-order outcomes, Table \ref{tab:case_study_comparison} details a cross-disciplinary task involving a "rote learner" with fragmented knowledge.
The baseline ChatGPT-EDU acts as an "Efficient Lecturer." In Round 2, it "spoon-feeds" the critical link between spores and humidity. While efficient, this bypasses cognitive struggle, resulting in passive acceptance ("Okay, I remember now") and superficial integration.
In contrast, ERL4SIIP functions as a "Strategic Socratic Guide," prioritizing activation. It fosters Critical Thinking by prompting physical observation ("turn the leaf over") to confront misconceptions (Round 1) and scaffolds Knowledge Transfer by explicitly withholding answers, forcing the student to map biological needs ("spores need water") to geographical evidence ("tap and shade") (Round 3). This engineered "productive struggle" culminates in a self-generated "Aha Moment" (Round 4), qualitatively validating the superior quantitative metrics by demonstrating how the agent compels cross-domain reasoning rather than mere information retrieval.

\section{Additional Analysis Details}

\subsection{Computational Efficiency Analysis}
\label{app:efficiency}

We validate the computational feasibility of the LoRA-Division Based architecture. As shown in Table \ref{tab:efficiency_app1}, evolving full parameters is prohibitive. By projecting the search space into low-rank manifolds, our approach reduces peak VRAM usage by approximately 4x compared to theoretical full-parameter requirements and significantly cuts training time compared to standard single LoRA evolution, enabling scalable training on standard hardware.











\end{document}